# Multiparty Quantum Remote Control

Yu-Ting Chen and Tzonelih Hwang

**Abstract**

This paper proposes a multiparty quantum remote control (MQRC) protocol, which allows several controllers to perform remote operations independently on a target state based on a shared entanglement of Greenberger-Home-Zeilinger (GHZ) state.

## 1. Introduction

The property of quantum entanglement has become a crucial ingredient in quantum information. Many correlated researches have been proposed successively. In particular, quantum teleportation, first introduced by Bennett *et al*. [1], allowing Alice to send an arbitrary single photon $|\psi\rangle = (\alpha|0\rangle + \beta|1\rangle)$ to a distant Bob via a classical channel, gains much more attention. In addition to teleportation, quantum entanglement also allows one to teleport an arbitrary unitary operation from Alice to Bob to operate on a quantum state. This manner is called "quantum remote control".

The first quantum remote control (QRC), proposed by Huelga *et al*. in 2001 [2], explores the implementation of an arbitrary unitary operation upon a distant quantum system. That is, Alice can apply an arbitrary unitary operation $U$ on the quantum state $|\psi\rangle$ held by Bob located in a distant place. This task can be completed by bidirectional quantum state teleportation (BQST). Afterwards, many related QRC protocols [3-5] have been proposed.

So far, these remote control protocols investigate only the case of a single controller. That is, only one user can perform a remote operation on the target quantum state. Consider the following scenario where there is a safe deposit box whose switch is controlled by the angle of the quantum state $|\psi\rangle$. Controllers, located in distant locations, can perform rotation operations on the switch of this box by means of remote controls. Only when controllers perform operations with a correct sum of angles, can the box be opened successfully. The multiparty quantum remote control (MQRC) could be very useful in this environment. Furthermore, in many applications in quantum cryptography, such as in quantum secret sharing (QSS) [6-8] or in controlled quantum secure direct communication (CQSDC) [9, 10] and etc., it also requires that several users perform operations independently on the target quantum state. Hence, this paper intends to explore quantum remote control over multi-controllers that could be useful in the above mentioned scenarios. Though Chen *et al.* in [11] also mentioned remote control over multi-controllers (called observers there), the operations performed by the controllers have to be negotiated among these controllers beforehand. On the contrary, the newly proposed scheme allows users to independently perform operations in the set $U = \{U_0(\theta), U_1(\theta)\}$ [3],

$$U_0(\theta) = \begin{pmatrix} e^{i\theta} & 0 \\ 0 & e^{-i\theta} \end{pmatrix}, U_1(\theta) = \begin{pmatrix} 0 & e^{i\theta} \\ -e^{-i\theta} & 0 \end{pmatrix}, \tag{1}$$

where $\theta$ is an arbitrary real parameter.

The rest of this paper is organized as follows. The protocol with two controllers is first discussed in Section 2. Then, Section 3 extends it to an N-controller scenario. Finally, Section 4 concludes our discussion.

## 2. Proposed QRC Protocol with Two Controllers

Suppose there are three participants, Alice, Charlie, the controllers, and Bob, the target quantum holder, in the system sharing a Greenberger-Home-Zeilinger (GHZ) state,

$$|\phi\rangle_{q_a q_b q_c} = \frac{1}{\sqrt{2}}(|000\rangle + |111\rangle)_{q_a q_b q_c}, \qquad (2)$$

where the 1$^{st}$ particle '$q_a$' is possessed by Alice, '$q_b$' by Bob, and '$q_c$' by Charlie. Bob has an arbitrary quantum state, also called the target state:

$$|\psi\rangle_B = (\alpha|0\rangle + \beta|1\rangle)_B, \qquad (3)$$

where $\alpha^2 + \beta^2 = 1$. The proposed protocol, enabling Alice and Charlie to execute $U_0(\theta)$ or $U_1(\theta)$ on $|\psi\rangle_B$, is described step by step as follows:

**Step1**  Bob first performs a controlled-NOT (CNOT) operation on his qubit pair $(q_b, B)$ with $q_b$ as a control qubit and $B$ as a target qubit. Then, Bob measures the qubit $B$ in Z basis $\{|0\rangle, |1\rangle\}$ and the composite quantum state becomes

$$\text{CNOT}_{q_b B}(|\phi\rangle_{q_a q_b q_c} \otimes |\psi\rangle_B)$$

$$= \text{CNOT}_{q_b B}\left[\frac{1}{\sqrt{2}}(\alpha|0000\rangle + \alpha|1110\rangle + \beta|0001\rangle + \beta|1111\rangle)_{q_a q_b q_c B}\right]$$

$$= \frac{1}{\sqrt{2}}[(\alpha|000\rangle + \beta|111\rangle)_{q_a q_b q_c}|0\rangle_B + (\alpha|111\rangle + \beta|000\rangle)_{q_a q_b q_c}|1\rangle_B]. \qquad (4)$$

Bob broadcasts the result of the measurement, $MR_B$, via a classical channel.

**Step2**  Alice performs a unitary operation $U_A(\theta_a)$ on her qubit $q_a$ as follows. If $MR_B$ is $|0\rangle$, then Alice performs $U_A(\theta_a)$ on $q_a$. Otherwise, she performs $U_A(-\theta_a)$ on $q_a$. Then, Alice measures $q_a$ in X basis $\{|+\rangle = \frac{1}{\sqrt{2}}(|0\rangle + |1\rangle), |-\rangle = \frac{1}{\sqrt{2}}(|0\rangle - |1\rangle)\}$ and sends her measurement result, $MR_A$, to Bob. Moreover, Alice has to inform Charlie which operation in $U_A$ she has done. That is, if $U_A = U_0$ ($U_A = U_1$), then $C_a = 0$ ($C_a = 1$). The message $C_a$ is sent to Charlie via a classical channel.

**Step3**  Before Charlie performs the unitary operation $U_C(\theta_c)$ on his qubit $q_c$, he would flip $q_c$ according to $C_a$. If it is 0, he does nothing, otherwise he applies $\sigma_x = |0\rangle\langle 1| + |1\rangle\langle 0|$ on $q_c$. Then, Charlie performs unitary operation $U_C(\theta_c)$ on $q_c$ and measures it just like Alice did. After that, Charlie also sends his measurement result, $MR_C$, to Bob via a classical channel.

**Step4** Finally, Bob can correct the state of the $q_b$ by performing a corresponding unitary operation $U_b \in \{I, \sigma_z, \sigma_x, i\sigma_y\}$ on $q_b$ (shown in Table 1),

$$I = |0\rangle\langle 0| + |1\rangle\langle 1|,$$

$$\sigma_z = |0\rangle\langle 0| - |1\rangle\langle 1|,$$

$$\sigma_x = |0\rangle\langle 1| + |1\rangle\langle 0|,$$

$$i\sigma_y = |0\rangle\langle 1| - |1\rangle\langle 0|, \tag{5}$$

which is the result state after Alice and Charlie have performed their unitary operations.

Table 1: Measurement results and the corresponding operations

| $U_A(\theta_a)$ | $U_C(\theta_c)$ | $MR_A$ | $MR_C$ | $U_b$ |
|---|---|---|---|---|
| $U_0(\theta_a)$ | $U_0(\theta_c)$ | $|+\rangle$ | $|+\rangle$ | $I$ |
| | | $|+\rangle$ | $|-\rangle$ | $\sigma_z$ |
| | | $|-\rangle$ | $|+\rangle$ | $\sigma_z$ |
| | | $|-\rangle$ | $|-\rangle$ | $I$ |
| $U_0(\theta_a)$ | $U_1(\theta_c)$ | $|+\rangle$ | $|+\rangle$ | $\sigma_x$ |
| | | $|+\rangle$ | $|-\rangle$ | $i\sigma_y$ |
| | | $|-\rangle$ | $|+\rangle$ | $i\sigma_y$ |
| | | $|-\rangle$ | $|-\rangle$ | $\sigma_x$ |
| $U_1(\theta_a)$ | $U_0(\theta_c)$ | $|+\rangle$ | $|+\rangle$ | $\sigma_x$ |
| | | $|+\rangle$ | $|-\rangle$ | $i\sigma_y$ |
| | | $|-\rangle$ | $|+\rangle$ | $i\sigma_y$ |

|           |           | $|-\rangle$ | $|-\rangle$ | $\sigma_x$ |
|-----------|-----------|-------------|-------------|------------|
|           |           | $|+\rangle$ | $|+\rangle$ | $I$        |
| $U_1(\theta_a)$ | $U_1(\theta_c)$ | $|+\rangle$ | $|-\rangle$ | $\sigma_z$ |
|           |           | $|-\rangle$ | $|+\rangle$ | $\sigma_z$ |
|           |           | $|-\rangle$ | $|-\rangle$ | $I$        |

It should be noted that when one performs $U_1(\theta)$ on the target state, the quantum state will be changed from $|0\rangle$ to $|1\rangle$ (or from $|1\rangle$ to $|0\rangle$). Because Alice and Charlie perform their unitary operations on distinct qubits which are entangled, Alice needs to send the classical message $C_a$ to Charlie in Step 2 to indicate which operation she has performed and to let Charlie adjust his state to the same one.

For example, suppose Alice wants to perform $U_1(\theta_a)$ and Charlie wants to perform $U_0(\theta_c)$ on the particle $B$, held by Bob. The final state of $|\psi\rangle_B$ should become

$$U_0(\theta_c)U_1(\theta_a)|\psi\rangle_B = \left(-\alpha e^{-i(\theta_a+\theta_c)}|0\rangle + \beta e^{i(\theta_a+\theta_c)}|1\rangle\right)_B. \tag{6}$$

Let $MR_B$ be $|0\rangle$ in Step 2. Alice performs $U_1(\theta_a)$ on $q_a$ and measures it in X basis. The state of the composite quantum system transforms to

$$U_1(\theta_a)(\alpha|000\rangle + \beta|111\rangle)_{q_a q_b q_c} = -\alpha e^{-i\theta_a}|1\rangle_{q_a}|00\rangle_{q_b q_c} + \beta e^{i\theta_a}|0\rangle_{q_a}|11\rangle_{q_b q_c}$$

$$= \tfrac{1}{\sqrt{2}}\Big[|+\rangle_{q_a}\left(-\alpha e^{-i\theta_a}|00\rangle + \beta e^{i\theta_a}|11\rangle\right)_{q_b q_c}$$

$$+|-\rangle_{q_a}\left(\alpha e^{-i\theta_a}|00\rangle + \beta e^{i\theta_a}|11\rangle\right)_{q_b q_c}\Big]. \tag{7}$$

Suppose $MR_A$ is $|+\rangle$, and because $C_a = 1$, Charlie has to apply $\sigma_x$ on $q_c$ first and then performs $U_0(\theta_c)$ on it in Step 3. The result state is given by

$$U_0(\theta_c)\sigma_x\left(-\alpha e^{-i\theta_a}|00\rangle + \beta e^{i\theta_a}|11\rangle\right)_{q_b q_c}$$

$$= U_0(\theta_c)\left(-\alpha e^{-i\theta_a}|01\rangle + \beta e^{i\theta_a}|10\rangle\right)_{q_b q_c}$$

$$= -\alpha e^{-i(\theta_a+\theta_c)}|0\rangle_{q_b}|1\rangle_{q_c} + \beta e^{i(\theta_a+\theta_c)}|1\rangle_{q_b}|0\rangle_{q_c}$$

$$= \tfrac{1}{\sqrt{2}}\Big[\left(-\alpha e^{-i(\theta_a+\theta_c)}|0\rangle + \beta e^{i(\theta_a+\theta_c)}|1\rangle\right)_{q_b}|+\rangle_{q_c}$$

$$+\left(\alpha e^{-i(\theta_a+\theta_c)}|0\rangle + \beta e^{i(\theta_a+\theta_c)}|1\rangle\right)_{q_b}|-\rangle_{q_c}\Big]. \tag{8}$$

If Charlie measures $q_c$ in X basis and gets the $MR_C$ is $|+\rangle$, Bob can perform a correct operation $\sigma_x$ to obtain the result state in Eq.(6).

In the above-mentioned scenario, Alice and Charlie perform only one unitary operation on the target state respectively. In fact, this protocol also allows controllers to perform multiple remote operations on the target quantum state independently. Alice may perform multiple unitary operations $U_A^1(\theta_{a_1})$, $U_A^2(\theta_{a_2}),\ldots,U_A^n(\theta_{a_n})$ on $q_a$ and measures it in X basis to get $MR_A$ in Step2, where $U_A^i(\theta_{a_i}) \in U, i = 1$ to $n$. We let $C_{a_i} = 0$ ($C_{a_i} = 1$) if $U_A^i = U_0$ ($U_A^i = U_1$). Alice has to calculate $C_A = C_{a_1} \oplus C_{a_2} \oplus \cdots \oplus C_{a_n}$, where $\oplus$ is a bitwise exclusive-OR operation, and then send it to Charlie via a classical channel. According to $C_A$, Charlie can decide whether he applies $\sigma_x$ on $q_c$ in Step3 or not. After Charlie performs multiple unitary operations $U_C^1(\theta_{c_1})$, $U_C^2(\theta_{c_2}),\ldots,U_C^m(\theta_{c_m})$ on $q_c$, where $U_C^j(\theta_{c_j}) \in U, j = 1$ to $m$, and measures it in X basis to get $MR_C$, he also calculates $C_c = C_{c_1} \oplus C_{c_2} \oplus \cdots \oplus C_{c_m}$, $C_{AC} = C_A \oplus C_c$ and sends it together with $MR_C$ to Bob in Step 3. Based on these Bob can determine the correct operation $U_b$ (shown in Table 2).

Table 2: Calculation results and the corresponding operations

| $C_{AC}$ | $MR_A \oplus MR_C$ | $U_b$ |
|---|---|---|
| 0 | 0 | $I$ |

| 0 | 1 | $\sigma_z$ |
| --- | --- | --- |
| 1 | 0 | $\sigma_x$ |
| 1 | 1 | $i\sigma_y$ |

## 3. Proposed Multiparty QRC Protocol

This section extends the proposed protocol to a multi-controller scenario with N controllers Alice$_1$, Alice$_2$, …, Alice$_N$ and Bob with a target quantum state $|\psi\rangle_B$. Suppose they share an (N+1)-particle GHZ state in the system,

$$|\psi\rangle = \frac{1}{\sqrt{2}}\left(|0\rangle_{q_{a_1}}|0\rangle_{q_{a_2}}\ldots|0\rangle_{q_{a_N}}|0\rangle_{q_b} + |1\rangle_{q_{a_1}}|1\rangle_{q_{a_2}}\ldots|1\rangle_{q_{a_N}}|1\rangle_{q_b}\right), \qquad (9)$$

where the particles $q_{a_1}, q_{a_2}, \ldots, q_{a_N}$ and $q_b$ are possessed by Alice$_1$, Alice$_2$, …, Alice$_N$ and Bob respectively.

**Step1**   Bob first performs a controlled-NOT operation on his qubit pair $(q_b, B)$ with $q_b$ as a control qubit and $B$ as a target qubit. Then, Bob measures the qubit $B$ in Z basis $\{|0\rangle, |1\rangle\}$ and broadcasts the result of the measurement, $MR_B$, via a classical channel.

**Step2**   Alice$_1$ performs a unitary operation $U_{A_1}(\theta_{A_1})$ on her qubit $q_{a_1}$ as follows. If $MR_B$ is $|0\rangle$, Alice$_1$ performs $U_{A_1}(\theta_{A_1})$ on $q_{a_1}$. Otherwise, she performs $U_{A_1}(-\theta_{A_1})$ on $q_{a_1}$. Then, Alice$_1$ measures $q_{a_1}$ in X basis $\{|+\rangle, |-\rangle\}$ and sends the result of measurement, $MR_{A_1}$, to Bob. Like the above protocol, Alice$_1$ has to send $C_{A_1}$ to Alice$_2$ via a classical channel.

**Step3**   Alice$_2$, Alice$_3$, …, Alice$_N$ perform their unitary operations on their possessed particles one by one like Alice$_1$ did.

**Step4**   According to $C_{A_1}, C_{A_2}, \ldots, C_{A_N}$ and $MR_{A_1}, MR_{A_2}, \ldots, MR_{A_N}$, Bob has to perform a corresponding unitary operation $U_b$ on $q_b$. Table 2 can be extended in a straightforward way for this purpose.

Similarly, Alice$_1$, Alice$_2$, ..., Alice$_N$ can also perform multiple remote operations subsequently and independently on the target state. With the classical information, $C_{A_1}, C_{A_2}, ..., C_{A_N}$ and $MR_{A_1}, MR_{A_2}, ..., MR_{A_N}$, Bob can recover the final state successfully.

## 4. Conclusions

This paper proposes a multiparty QRC protocol using the GHZ state. Based on the property of entanglement, each controller can perform unitary operations from the set $U$ independently on a target quantum state. Because the angles $\theta'$s in the unitary operations are determined by controllers, Bob has no knowledge about this information. However, by means of classical communications, Bob still can obtain the result quantum state.

## Acknowledgement

We would like to thank the National Science Council of Republic of China for financial support of this research under Contract No. NSC 100-2221-E-006-152-MY3 and No. NSC 101-2221-E-006-266-.